\documentclass[usegraphicx,usenatbib,doublespace]{mn2e}
\usepackage{amsmath}
\usepackage{amssymb}
\usepackage{bm}
\usepackage{epsfig}

\begin{document}
\title{Ionized bubble number count as a probe of non-Gaussianity}

\author[Tashiro, H. et al.]
{Hiroyuki Tashiro$^1$, and Naoshi Sugiyama$^{2,3,4}$\\
$^1$Center for Particle Physics and Phenomenology (CP3), 
Universit\'e catholique de Louvain,\\
Chemin du Cyclotron, 2, B-1348 Louvain-la-Neuve, Belgium\\
$^2$Department of Physics and Astrophysics, Nagoya University, 
  Chikusa, Nagoya 464-8602, Japan\\
$^3$Institute for Physics and Mathematics of the Universe, 
University of Tokyo,\\ 
 5-1-5 Kashiwa-no-Ha, Kashiwa,
Chiba, 277-8582, Japan \\
$^4$Kobayashi-Maskawa Institute for the Origin of Particles and the Universe,
Nagoya University, Nagoya 464-8602, Japan }
\date{\today}

\maketitle

\begin{abstract}

The number count of ionized bubbles on a map of 21~cm fluctuations
with the primordial non-Gaussianity is investigated.  The existence of
the primordial non-Gaussianity modifies the reionization process,
because the formation of collapsed objects, which could be the source
of reionization photons, is affected by the primordial
non-Gaussianity.  In this paper, the abundance of ionized bubbles is
calculated by using a simple analytic model with the local type of the
primordial non-Gaussianity, which is parameterized by $f_{NL}$.  In
order to take into account the dependence of the number count on the
size of ionized bubbles and the resolution of the observation
instrument, a threshold parameter $B_b$ which is related to the the
surface brightness temperature contrast of an ionized bubble is
introduced.  We show the potential to put the constraint on $f_{NL}$ 
from the number count by future observations such as LOFAR and SKA.

\end{abstract}

\section{introduction}

Recent cosmic microwave background (CMB) observations have revealed
the statistical nature of primordial fluctuations and given the
strong support of the inflationary scenario 
\citep{komatsu-wmap-2009}.  However,  it is still
difficult to specify the model among the inflationary scenario from CMB observations. 
The measurement (or even the upper bound) of non-Gaussianity of primordial fluctuations is, 
on the other hand, expected to have the potential to rule out many inflationary models. Although 
the primordial fluctuations are predicted as nearly Gaussian 
in all inflationary models, the degree
of the deviation from the Gaussianity depends on each specific  model.
The slow-roll inflation models with a single scalar field produce
almost pure Gaussian fluctuations 
\citep{guth-pi-1982, starobinsky-1982,bardeen-steinhardt-1983}, 
and the deviation from the Gaussianity is unobservably small
\citep{falk-rangarajan-1993, gangui-lucchin-1994}. 
Meanwhile some multi-field inflation scenarios can produce large non-Gaussianity 
which can be observed in the near future experiments such as PLANCK mission
\citep{battefeld-easther-2007}. For comprehensive review see 
\citet{bartolo-komatsu-2004}.

The most stringent constraint on the primordial non-Gaussianity is 
currently obtained from CMB measurements \citep{verde-wang-2000}. 
Wilkinson Microwave Anisotropy Probe (WMAP) sets the limit 
on the so-called local type of the primordial non-Gaussianity,
which is parameterized by the constant dimensionless parameter
$f_{NL}$. From WMAP 5yr data, the constraint was
$-9<f_{NL}<111$ \citep{komatsu-wmap-2009}. 
Recently, this constraint is updated by 
\citet{curto-martinezgonzales-2009}, $-18<f_{NL}<80$.

The large scale structure and the abundance of the collapsed objects,
e.g. galaxy clusters and galaxies, are alternative probes of the 
non-Gaussianity. The positive $f_{NL}$ enhances the abundance of the
collapse objects, while the negative $f_{NL}$ decreases the abundance.
In particular, the formation process of rare objects such as high-mass collapsed objects at high 
redshift, which is controlled by the high density tail, is quite sensitive to $f_{NL}$ 
(see \citealt{desjacques-seljak-2010} for a review).  On the 
contrary, the abundance of a void, which is formed in the underdense region, 
is sensitive to the low density tail of the
fluctuation distribution, and has also the potential to prove the
primordial non-Gaussianity 
\citep{grossi-branchini-2008,kamionkowski-verde-2009}.

The non-Gaussianity also affects the reionization process, because
the formation of high-mass collapsed objects, which could be the 
source of reionization photons, is enhanced
(reduced) by positive (negative) $f_{NL}$ 
\citep{chen-cooray-2003, crociani-moscardini-2009}.
Therefore, in this paper, we study the potential of the number
count of ionized bubbles on a map of the redshifted 21~cm
line fluctuations from neutral hydrogen as a probe of the 
non-Gaussianity. 
Now, 
LOFAR\footnote{http://www.lofar.org}, MWA\footnote{http://www.mwatelescope.org/}
and SKA\footnote{http://www.skatelescope.org} are being installed or designed for the measurements of
21~cm line fluctuations.
The maps of 21~cm fluctuations are sensitive to the density, temperature, and ionized fraction of the intergalactic medium (IGM).
Studying the 21-cm tomography tells us about
the physics of IGM gas and structure formation during the epoch of
reionization.
\citep{madau-meiksin-rees-1997,
tozzi-madau-2000,ciadri-madau-2003,furlanetto-zaldarriaga-2004}.
An ionized bubble will be observed as a dark 
spot on a map of 21~cm fluctuations,
because hydrogens are fully ionized in the inside of the bubble.
We employ a simple analytic model of the ionized bubble to 
investigate the non-Gaussian effect on the number count of 
ionized bubbles.

The outline of this paper is the following.
In Sec.~II, we give the simple analytic model of the ionized bubble
based on the Press-Schechter theory with the non-Gaussianity $f_{NL}$.
In Sec.~III, we calculate the surface brightness temperature of a
ionized bubble and show the results of the number count of ionized 
bubbles. Section IV is devoted to the conclusion.  
Throughout the paper, we use the concordance cosmological parameters for a
flat cosmological model, i.e. $h=0.73 \ (H_0=h
\times 100 {\rm ~km/s / Mpc})$, $T_0 = 2.725$K, $\Omega _{\rm b}
=0.05$, $\Omega_{\rm m} =0.27$ and $\sigma_8=0.8$.

\section{analytic model of ionized bubbles}

In this paper, we focus on the so-called local type of the
primordial non-Gaussianity, which is parameterized by the constant
dimensionless parameter $f_{NL}$ as \citep{komatsu-spergel-2001}
\begin{equation}
\Phi({\bm x}) = \Phi_{\rm G}({\bm x}) +f_{NL}(\Phi_{\rm G}^2({\bm x}) 
- \langle \Phi_{\rm G}^2 ({\bm x})\rangle), 
\end{equation}
where $\Phi$ is Bardeeen's gauge-invariant potential,
$\Phi_{\rm G}$ is the Gaussian part of the potential and
$\langle ~ \rangle$ denotes the ensemble average.
The power spectrum of $\Phi_{\rm G}$ is defined by
\begin{equation}
\langle \Phi_{\rm G} ({\bm k}) \Phi_{\rm G} ({\bm k}') \rangle
=(2 \pi)^3 \delta_D({\bm k} + {\bm k}' ) P(k),
\end{equation}
where $\delta_D$ is Dirac's delta function.
The bispectrum of the local type potential can be written as
\begin{equation}
\langle \Phi ({\bm k}_1) \Phi ({\bm k}_2) \Phi ({\bm k}_3)\rangle =
(2\pi)^3 \delta_D({\bm k}_1 + {\bm k}_2 + {\bm k}_3) B({\bm k}_1,{\bm k}_2,{\bm k}_3),
\end{equation}
where
$B({\bm k}_1,{\bm k}_2,{\bm k}_3)$ can be written with $f_{NL}$ and
the power spectrum $P(k)$,
\begin{equation}
B({\bm k}_1,{\bm k}_2,{\bm k}_3) = 2 f_{NL} (P({\bm k}_1) P({\bm k}_2) +P({\bm k}_1) P({\bm k}_3)
+P({\bm k}_2) P({\bm k}_3)).
\end{equation}

The existence of the primordial non-Gaussianity modifies the mass
function obtained under the assumption of the Gaussian density 
fluctuation. The mass function for the non-Gaussian case is given
by estimating the fractional correction to the one for the Gaussian
case in the Press-Schechter theory 
\citep{matarrese-verde-2000,loverde-miller-2008},
\begin{equation}
\frac{dn(M, z)}{dM}=
-\sqrt{\frac{2}{\pi}}\frac{\bar{\rho}}{M} \exp \left[{-\frac{\delta^2_c}{2\sigma_M^2}}\right]
 {\cal R}_{NG},
\end{equation}
where ${\cal R}_{NG}$ is the non-Gaussian correction. We adopt
the correction of \citet{loverde-miller-2008},
\begin{eqnarray}
{\cal R}_{NG}&=&
\left[\frac{d\textrm{ln}\sigma_M}{dM}\left(\frac{\delta_c}{\sigma_M}+\frac{S_3\sigma_M}{6}\left(\frac{\delta_c^4}{\sigma_M^4}-2\frac{\delta_c^2}{\sigma_M^2}-1\right)\right)\right.
\nonumber \\
&& \qquad
\left. +\frac{1}{6}\frac{dS_3}{dM}\sigma_M
\left(\frac{\delta_c^2}{\sigma_M^2}-1\right)\right],
\label{eq:ng-correction}
\end{eqnarray}
where $\delta_c$ is the critical density contrast for collapsing
objects, $\sigma_M$ and $S_3$ are the smoothed density dispersion 
and the smoothed skewness at a redshift $z$ with a top-hat window 
function $W(R,k)$.
The smoothed density dispersion $\sigma_M(M)$ is given by
\begin{equation}
\sigma^2 _M(M)=\langle \delta_R^2 \rangle
=\int {dk^3\over (2 \pi)^3} W(R,k)^2 D(k,z)^2 P (k),
\end{equation}
where $R$ is the scale corresponding to $M$ in the top-hat window
function and $D(k,z)$ is the relation between $\Phi$ and $\delta$
in the linear perturbation, $\delta(k,z)=D(k,z) \Phi(k)$.
The smoothed skewness $S_3(M)$ is obtained by
\begin{eqnarray}
S_3(M) &\equiv& \frac{\langle\delta_R^3\rangle}
{\langle\delta_R^2\rangle^{2}},
\\
\langle\delta_R^3\rangle &=&
\int {d^3k_1 \over (2 \pi)^3}
{d^3k_2 \over (2 \pi)^3}
{d^3k_3 \over (2 \pi)^3} W(R,k_1) W(R,k_2) W(R,k_3) 
\nonumber \\
&& \times D(k_1,z) D(k_2,z) D(k_3,z)
\langle \Phi(k_1) \Phi(k_2) \Phi(k_3) \rangle.
\end{eqnarray}
The modification of the mass function from the existence of non-Gaussian fluctuations 
is described by the terms proportional to $S_3$.  

The reionization involves various physical processes such as
star formation, radiative transfer through IGM,
and so on. The aim of this paper is to demonstrate the potential of
the number count of ionized bubbles as a probe of the
non-Gaussianity.  Without discussing the details of the reionization
process, we employ a simple analytic reionization model.  First, we
assume that a halo whose virial temperature is larger than $10^4~$K
can collapse and the source of ionizing photons can be produced in the
inside of such collapsed halos.  Secondly, the shape of the ionized
bubble produced by a collapsed halo is spherical and the size depends
on the halo mass.  Under these assumptions, we give the ionizing
photons per baryon in a collapsed halo as
\begin{equation}
N_{\rm ion} = N_\gamma f_* f_{\rm esc},
\end{equation}
where $N_\gamma$ is the number of ionizing photons per baryon 
in stars, $f_*$ is the star formation efficiency, and $f_{\rm esc}$ 
is the escape
fraction of ionizing photons into the IGM. We adopt $f_*=0.1$ and 
$f_{\rm esc}=0.05$ in this paper. 
The parameter $N_\gamma$ depends on the ionizing source. Here
we consider population III stars as the ionizing sources and 
take $N_\gamma=44,000$ \citep{bromm-kudritzki-2001}.
Assuming that all ionizing photons ionize hydrogen atoms and
recombinations are negligible, 
we obtain the maximum physical size of an ionized bubble in the IGM
as \citep{loeb-barkana-2005}
\begin{eqnarray}
R_{\rm max} &=& 0.138\, f_{\rm esc}^{1/3}\, 
\left( \frac{M} {10^9\, M_{\odot}} \right)^
{1/3}\, \left( \frac{1+z} {11} \right)^{-1}
\nonumber \\
&& \quad \times
\left( \frac{\Omega_m h^2} {0.14} \right)^{-1/3}\,
\left( \frac{N_{\gamma} f_*} {430}
\right)^{1/3}\ {\rm Mpc}\ .
\label{eq:bubble-size}
\end{eqnarray}

\section{number count of ionized bubbles}

An ionized bubble is detected as a hole with radius $R_{\rm max}$
on a map of 21~cm fluctuations because the inside of an ionized 
bubble is assumed to be fully ionized. In an actual observation,
however the hole is smeared due to the finite beam size of a 
telescope with the signal from surrounding neutral hydrogens.
Taking into account this smearing effect, we define the surface
brightness temperature contrast of an ionized bubble as
\begin{equation}
B(\theta,z) = {Y_{obs}(\theta, z) \over T_{21}(z) \int d \Omega'~
   \exp \left[- {\theta'^2 \over 2 \sigma^2} \right]},
\end{equation}
where $\sigma=\theta_{\rm FWHM}/\sqrt{8 \ln 2}$ with 
$\theta_{\rm FWHM}$ being the FWHM resolution of the observation,
and $T_{21}(z)$ is the 21 cm background temperature at $z$.
The surface brightness temperature of an ionized bubble
$Y_{obs}(\theta)$ is obtained by
\begin{equation}
Y_{obs} (\theta,z) = \int d \Omega ' ~T(\theta-\theta',z )   
\exp \left[- {(\theta-\theta')^2 \over 2 \sigma^2} \right],
\label{eq:Y-def}
\end{equation}
where the angular profile of the ionized region $T(\theta, z )$ is 
given by
\begin{equation}
T(\theta,z )= \left \{
\begin{array}{c}
0,\quad \theta < \theta_R
\\
T_{21}(z),\quad \theta > \theta_R
\end{array}
\right . .
\label{eq:stepfunc}
\end{equation}
Here $\theta_R$ is the angular size corresponding to $R_{\rm max}$.

\begin{figure}
  \begin{center}
\includegraphics[keepaspectratio=true,height=60mm]{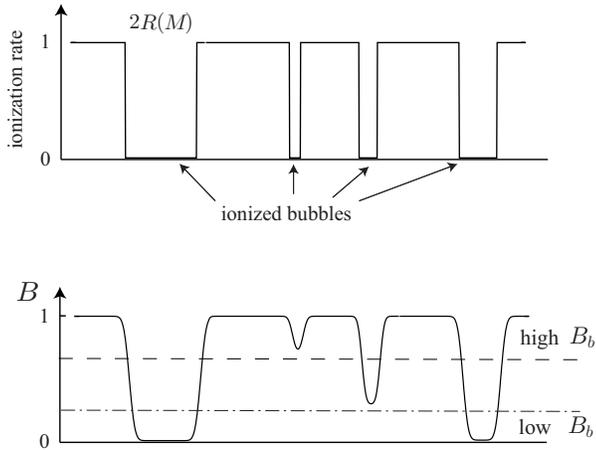}
  \end{center}
  \caption{The sketches of ionized bubbles and the smearing effect
by the finite observation resolution. 
Ionized bubbles are fully ionized in
the inside of each radius $R(M)$ (top panel). The finite resolution
smoothes the signal from ionized bubble. The smoothed surface
brightness temperature contrast $B$ for each bubble is different, depending on both the size of ionized bubbles and the resolution 
size (bottom panel). As a result, each threshold $B_b$ gives different number count of ionized bubbles.}
  \label{fig:bubble-fig}
\end{figure}

As the criterion for the detection of an ionized bubble,
we introduce a parameter $B_b$. We count bubbles with $B$ smaller than $B_b$ as the detectable ones on a 21~cm map 
(see also Fig.~\ref{fig:bubble-fig}).
Considering Eqs.~(\ref{eq:bubble-size}), (\ref{eq:Y-def}) 
and (\ref{eq:stepfunc}), 
we can relate $B_b$ to the limiting mass $M_{\rm lim}$
which is the mass of a collapsed halo associated to an ionized 
bubble with $B=B_b$.
Therefore, the number count of ionized bubbles with $B < B_{b}$ 
is given by
\begin{equation}
N(<B_b) = \int_{M_{\rm lim}(B_b)} dM {dV \over dz} {d n \over dM}
\Delta z,
\end{equation}
where $V$ is a comoving volume and $\Delta z$ is a width of the
observation redshift slice. 
We set $\Delta z=0.1$ throughout this paper.

\begin{figure}
  \begin{center}
\includegraphics[keepaspectratio=true,height=60mm]{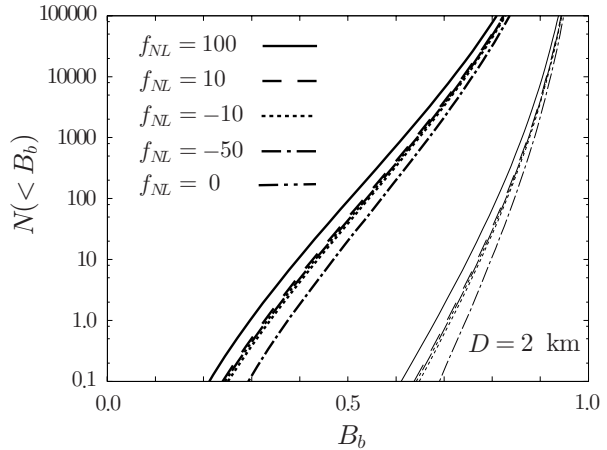}
  \end{center}
  \begin{center}
\includegraphics[keepaspectratio=true,height=60mm]{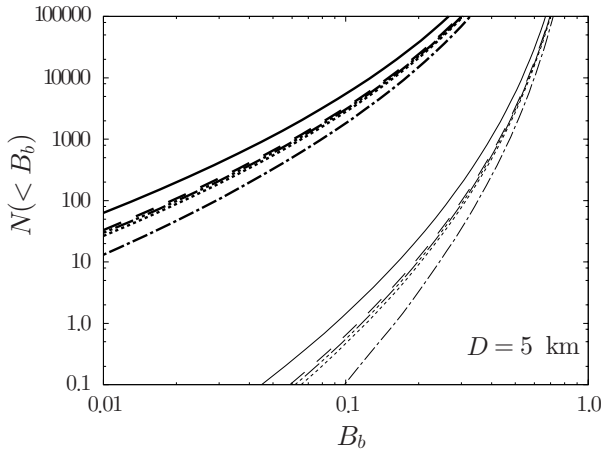}
  \end{center}
  \caption{
  The number count of ionized bubbles as a function of $B_b$
for different $f_{NL}$. We plot the number counts
  for $z=11$ and $z=13$ as the thick and thin lines, respectively. 
  We use the low resolution type ($D=2~$km) in the top panel and the high resolution type ($D=5~$km) in the bottom. }
  \label{fig:numbercount}
\end{figure}

In Fig.~\ref{fig:numbercount}, we plot $N(<B_b)$ as a function 
of $B_{b}$ for different $f_{NL}$. The 21~cm fluctuations 
during the epoch of reionization will be mapped by 
radio interferometric observations. The resolution 
$\theta_{\rm FWHM}$ for a wave length $\lambda$
depends on its baseline $D$ as $\theta_{\rm FWHM}= \lambda /D$. 
In the calculation of Fig.~\ref{fig:numbercount}, we choose 
two types of the baseline; low resolution type, $D=2~$km, which
corresponds to the LOFAR baseline, and high resolution type, 
$D=5~$km, which corresponds to the SKA baseline.

As is shown in Fig.~\ref{fig:numbercount}, the number count strongly depends on the resolution and the redshift
of the observation. The large angular resolution smears the ionized
bubble whose size is smaller than the resolution. Then,
the more the redshift decreases, the larger the number of halos which grow enough to collapse is. In addition, both physical and angular
size of a bubble at high redshifts is small compared to the one 
with the same halo mass at low redshifts. 
As a result, the number count strongly
decreases with increasing of the angular resolution or the redshift. 
The number count with $B_b=0.1$ for the high resolution type at $z=11$ is about
2000, while the number count for the low resolution type at $z=13$ is less 
than one even for full sky survey with $B_b=0.5$ as is shown in 
Fig.~\ref{fig:numbercount}.

\begin{figure}
  \begin{center}
\includegraphics[keepaspectratio=true,height=100mm]{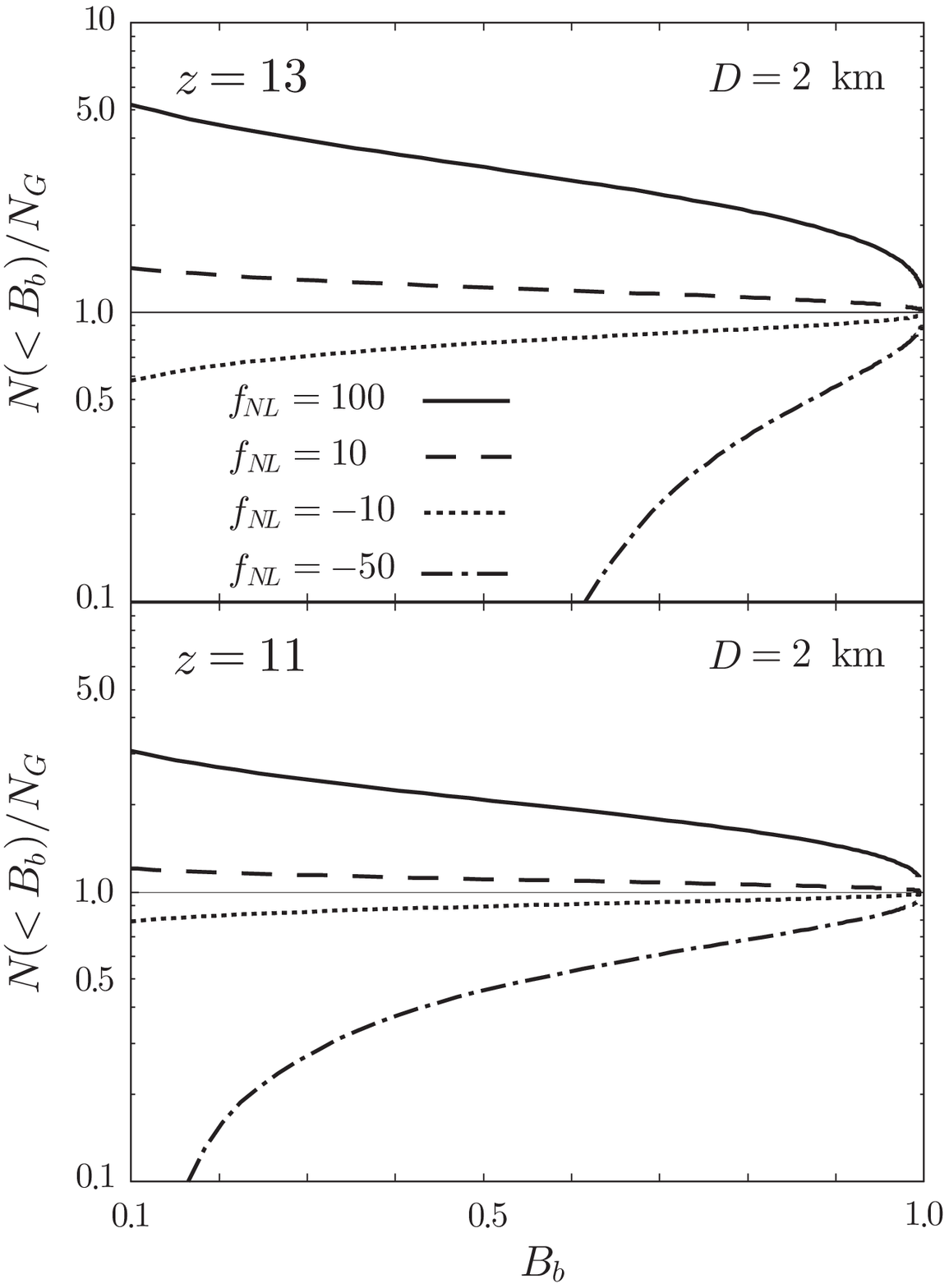}
  \end{center}
  \begin{center}
\includegraphics[keepaspectratio=true,height=100mm]{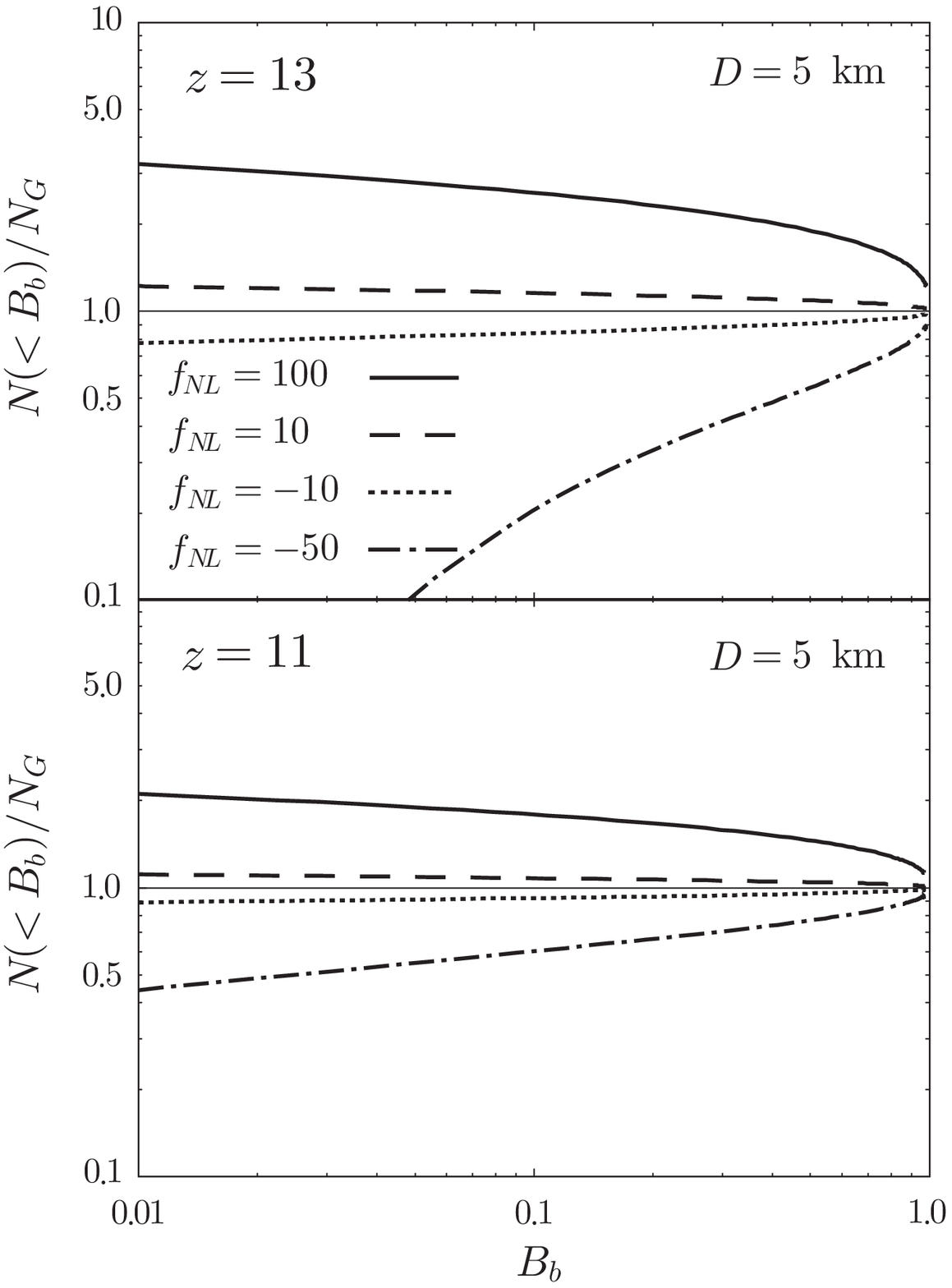}
  \end{center}
  \caption{
  The fractional deviation from the number count for the case of the
  Gaussian mass function as a function of $B_b$.  We use the low
 resolution type ($D=2~$km) and the high resolution type ($D=5~$km)
in the top two panels and the bottom two panels, respectively. In both angular resolution cases, we set $z=13$ and $z=11$.
The curves are for $f_{NL}=100$, $f_{NL}=10$, $f_{NL}=-10$
and $f_{NL}=-50$ from top to bottom in both panels.}
  \label{fig:contra-number1}
\end{figure}

To make clear the difference of the number counts between the 
Gaussian and non-Gaussian primordial fluctuations, we plot the 
ratio of the number counts between non-Gaussian and Gaussian cases
as a function of $B_b$ in Fig.~\ref{fig:contra-number1}. 
Producing ionized bubbles during the epoch of reionization is a 
rare event. Therefore
the number counts depend on the degree of the primordial 
non-Gaussianity. This
tendency is emphasized as the redshift increases. The 
deviation from the Gaussian case becomes larger for all 
non-Gaussian cases at $z=13$ than at $z=11$. 
The difference is also enhanced with decreasing 
$B_b$ because lower $B_b$ corresponds to higher threshold halo mass $M_{lim}$.  

It is shown in Fig.~\ref{fig:contra-number1} that the 
dependence on $f_{NL}$ of the number count is more significant for the negative value than the  positive one. In the case of negative 
$f_{NL}$,the high density tail of the distribution function is
suppressed, and less number of bubble formation, which is a rare
event, takes place.  Therefore with decreasing $B_b$, the number 
count rapidly decreases as is shown for $f_{NL}=-50$.  On the other
hand, for positive $f_{NL}$, the high density tail of the 
distribution function is enhanced.  Accordingly, there are more 
bubble formation for larger $f_{NL}$.  However, the terms 
proportional to $S_3$ and $dS_3/dM$ in Eq.~(\ref{eq:ng-correction})
have opposite sign in the case of positive $f_{NL}$.  Therefore the
dependence on $f_{NL}$ is not as strong as in the case of negative 
$f_{NL}$.  

By comparing upper two panels with lower two panels of 
Fig.~\ref{fig:contra-number1}, 
we find that deviation from the 
Gaussian case is larger in the lower resolution observation.   
However, number counts become smaller for the lower resolution case. 
As a result, the error of the determination on $f_{NL}$ is expected
to be large in the low resolution observation.
We discuss this point in the term of signal to noise ratio of the 
detection later.

\begin{figure}
  \begin{center}
\includegraphics[keepaspectratio=true,height=60mm]{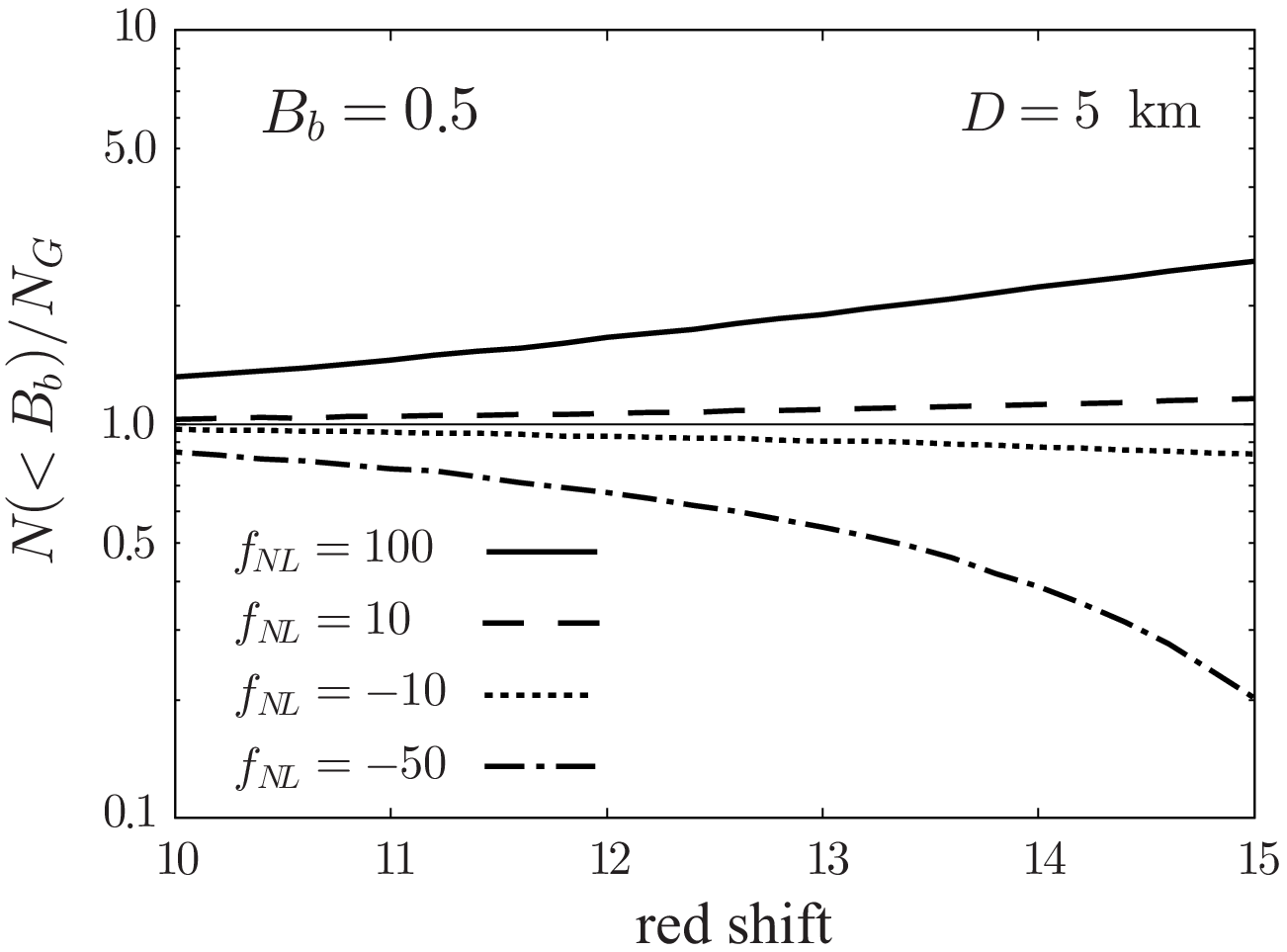}
  \end{center}
  \begin{center}
\includegraphics[keepaspectratio=true,height=60mm]{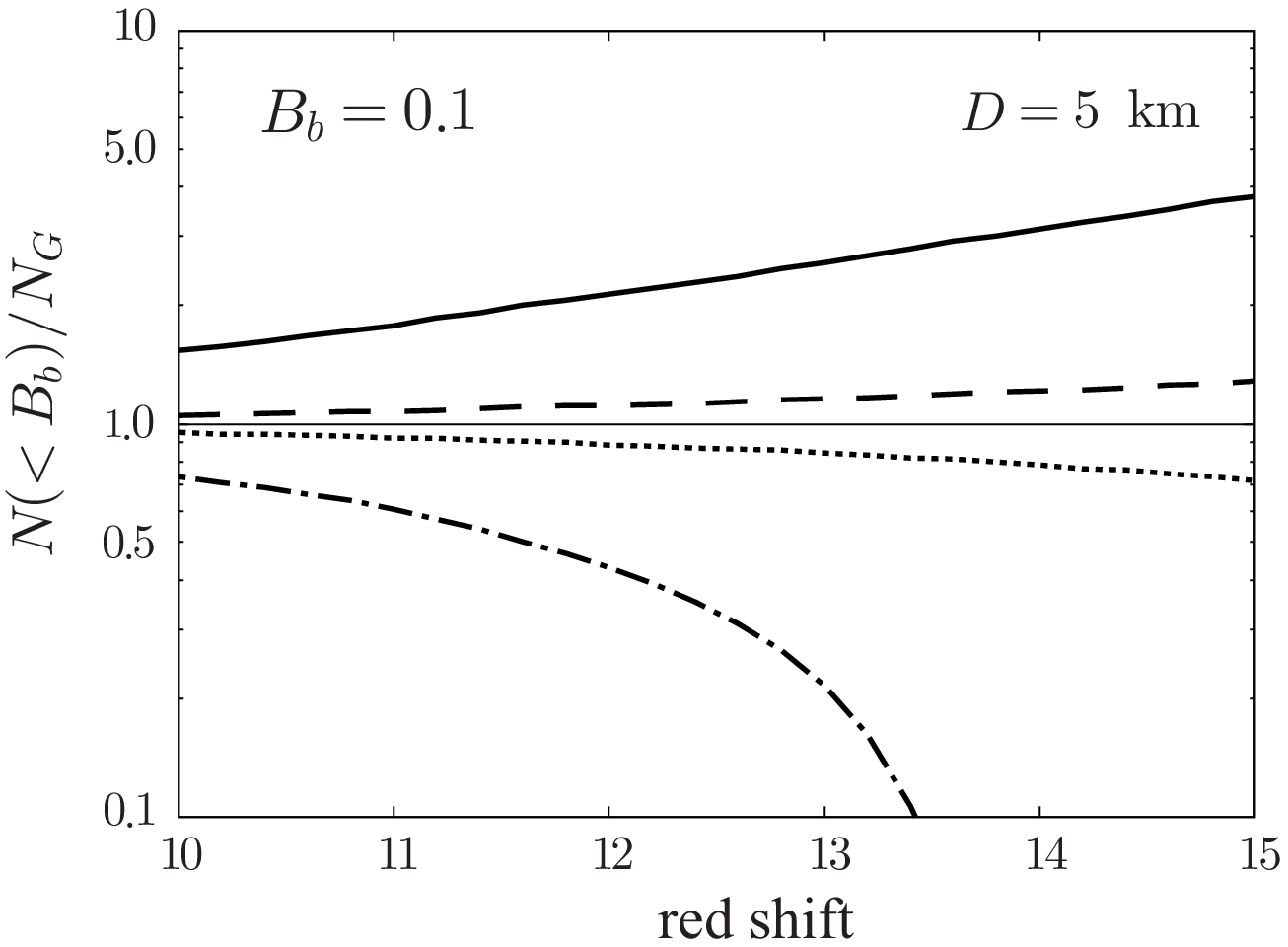}
  \end{center}
  \caption{The evolution of the fractional deviation from the number
  count for the case of the
  Gaussian mass function as a function of redshift. We set $B_b=0.5$ in the top panel and $B_b=0.1$ in the bottom panel.
The curves are for $f_{NL}=100$, $f_{NL}=10$, $f_{NL}=-10$
and $f_{NL}=-50$ from top to bottom in both panels.}
  \label{fig:contra-number}
\end{figure}

In Fig.~\ref{fig:contra-number}, we show the evolution of the
deviation from the Gaussian case with fixing $B_b$. 
In these figures, we adopt the high resolution type
and set $B_b=0.5$ and $0.1$. The figures tell us that increasing 
redshift or decreasing $B_b$ makes the deviation large. 
In particular, in high redshift or low $B_b$ cases,
the number count for the negative $f_{NL}$ case damps rapidly due to
the suppression of the bubble formation.

In order to discuss the measurement of $f_{NL}$ by the number
count of  bubbles, 
we calculate the Signal to Noise ratio. We assume
that the observation instrument is ideal, i.e., there is no instrumental noise, and the foreground noise can be completely removed. 
Therefore we take into account only the Poisson noise of the number
count. The SN ratio for fixed $B_b$ is given by
\begin{equation}
S/N= \sqrt{f_{sky} N(<B_b)},
\label{eq:SNratio}
\end{equation}
where $f_{sky}$ is the fraction of sky observed.  
According to Eq.~(\ref{eq:SNratio}), the large number count gives 
high SN ratio. Therefore, the larger
the threshold $B_b$ is set, the higher the SN ratio becomes. 

However, setting high threshold makes two problems arise.
The first problem is difficulty counting bubbles. 
In the large $B_b$ case, we need to count not only large bubbles 
but also small bubbles which are hardly seen in a 21~cm map. 
The second problem is that the deviation from the Gaussian case 
becomes smaller as $B_b$ increases 
as is shown in Fig.~\ref{fig:contra-number1}.
It should be notice that larger S/N does not necessarily mean 
sensitive to the non-Gaussianity detection.  
Therefore we need to optimize for the threshold value $B_b$
corresponding to the observational settings.

\begin{figure}
  \begin{center}
\includegraphics[keepaspectratio=true,height=60mm]{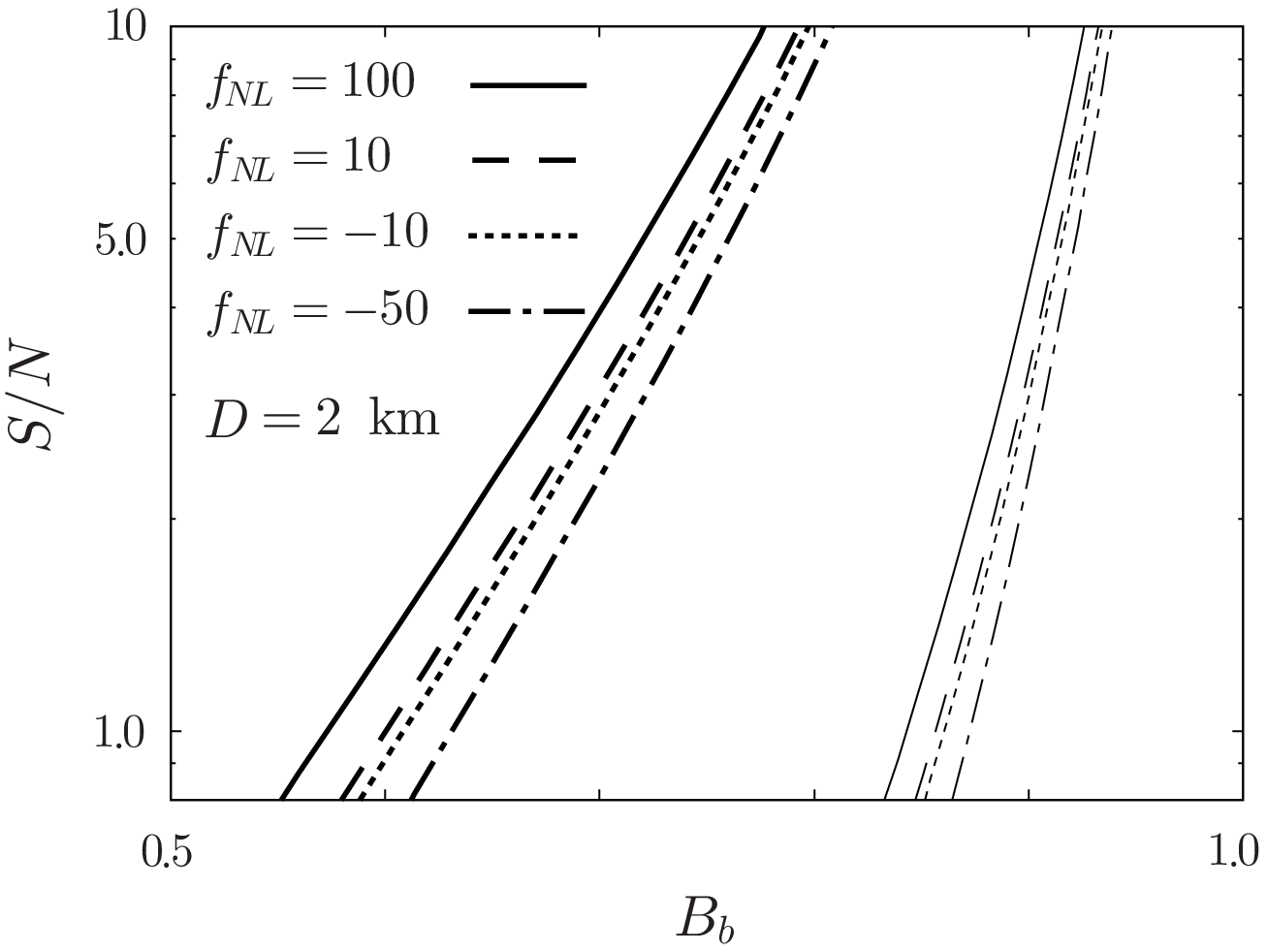}
  \end{center}
  \begin{center}
\includegraphics[keepaspectratio=true,height=60mm]{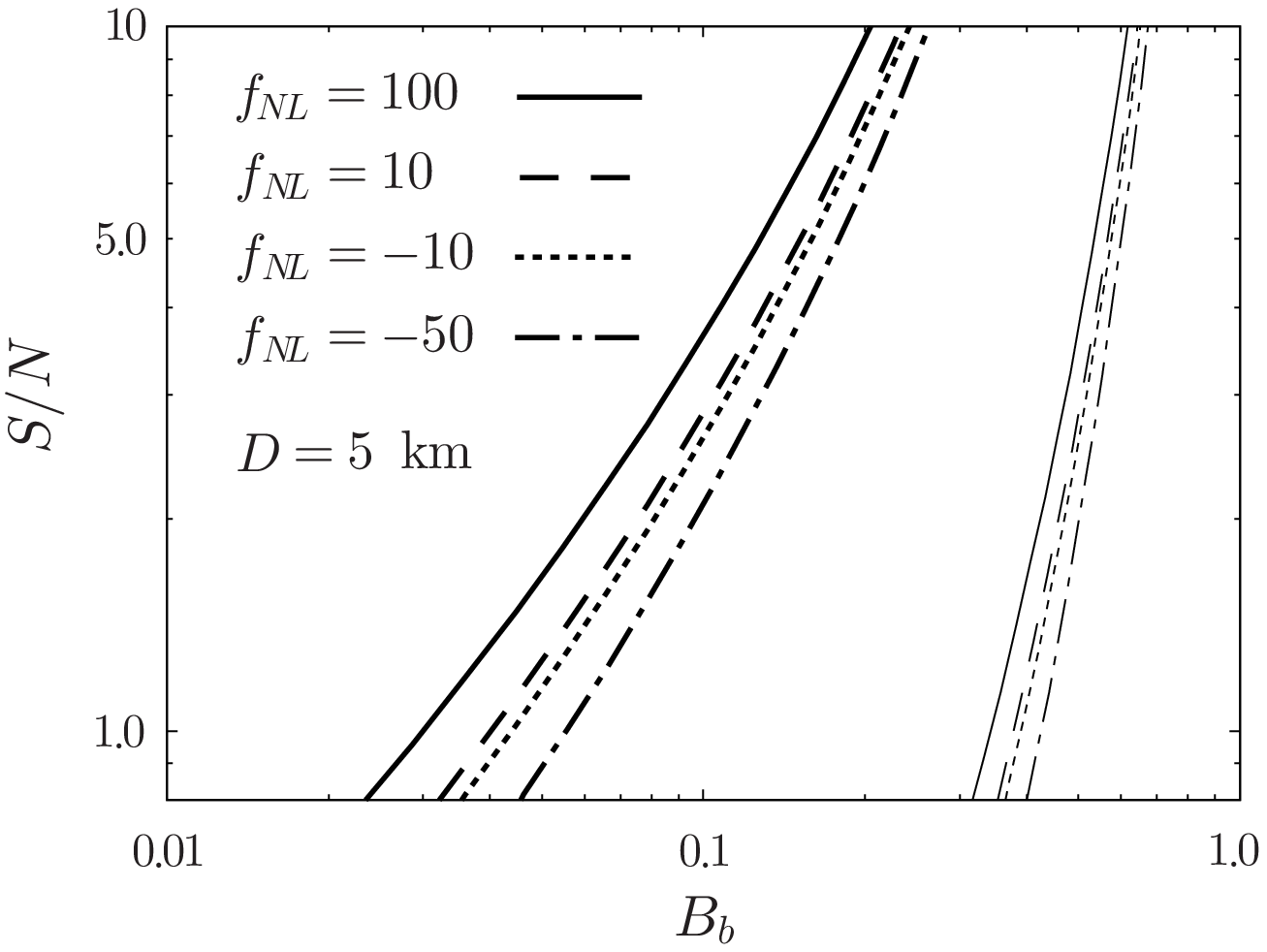}
  \end{center}
  \caption{The SN ratio of the number count as a function of $B_b$.
We set $D=2$~km in the top panel, $D=5$~km in the bottom panel. 
We assume $f_{sky}=2.42 \times 10^{-3}$.}
  \label{fig:snratio}
\end{figure}

In Fig.~\ref{fig:snratio},
we plot the SN ratio with
adopting SKA's field of view ($\sim 100~{\rm deg^2}$ for 140 MHz), 
which corresponds to $f_{sky}=2.42 \times 10^{-3}$.
The top panel of Fig.~\ref{fig:snratio} is for 
the low resolution ($D=2$~km) and the bottom one 
is for the high resolution ($D=5$~km).
In order to obtain $S/N>5$ for $f_{NL} =100$
for the low resolution at $z=11$
we must set $B_b > 0.7$ at least.
With $B_b \sim 0.7$, according to Fig.~\ref{fig:numbercount},
the number counts for $f_{NL}=100$ is 
$\sim 20$ for SKA's field of view, while the one for $f_{NL}=0$ are 
$\sim 10$ for SKA's field of view (it should be reminded that 
Fig.~\ref{fig:numbercount} shows the number counts for the full sky survey, $f_{sky}=1$).
On the other hand, in the high resolution, $B_b > 0.1$ is 
enough to obtain $S/N \sim 5$ for $f_{NL}=100$. In this case, 
the number count for $f_{NL}=100$ is $\sim 20$, while
the one for $f_{NL}= 0$ is $\sim 10$ with $S/N \sim 3$. 
Therefore we conclude that
the number count of ionized bubbles has potential to give 
the constraint on $f_{NL}$.

\section{conclusion}

In this paper, we have studied the number count of ionized bubbles on a 21~cm map as a probe of the primordial 
non-Gaussianity. 
Although we have used a simple analytic model of an ionized bubble,
such a simple model is enough to illustrate the relation 
between $f_{NL}$ and the number count of ionized bubbles. 

In order to take into account the dependence of the number count 
on the size of ionized bubbles and the resolution of the observation instrument, we have introduced a threshold parameter
$B_b$. The bubbles with $B<B_b$ are counted as the
detectable ionized bubbles on the 21~cm map. Setting smaller $B_b$
means counting larger size bubbles. 
Therefore, we can expect large numbers of ionized bubbles for large
$B_b$. The deviation of the number count for the non-Gaussian 
case from the one for the Gaussian case depends on the threshold
$B_b$. Small $B_b$ brings large deviation. However the number count
for small $B_b$ becomes small. As a result, 
the SN ratio goes to low. In order to constrain $f_{NL}$ by 
the number count of ionized bubbles, the optimization 
between $B_B$ and the resolution of observations is needed. 
For example, in our model, we have found that
the number counts for $f_{NL}=100$ and $f_{NL}=0$ for the high
resolution with $B_b=0.1$ are about $20$ with $S/N \sim5$ 
and about $10$ with $S/N \sim 3$, respectively.  

The number count also depends on the redshift of a 21~cm map.
Compared with the number counts at low redshifts, the ones at
high redshifts are lower sensitive to $f_{NL}$. However the number counts at high redshifts are relatively smaller than at low redshifts.Therefore, the SN ratio at high redshifts becomes large and
it becomes difficult to give the constraint on $f_{NL}$.
Besides, the detection of ionized bubble itself is a challenge
\citep{datta-bharadwaj-2007,datta-bharadwaj-2009}. 

In this paper, we have not taken into account the effect of overlapping of ionized bubbles.
The overlaps take place as the universe evolves. 
As is shown in many reionization simulation works, once overlapping
proceeds, the number count of ionized bubbles quickly 
decreases and the size of ionized bubbles becomes rapidly increasing.
Moreover the evolution of the overlapping process depends strongly 
on the model of reionization.
Therefore, our results in this paper can be only adapted 
for the early phase of the epoch of reionization before the overlapping process takes place. 

In order to properly treat the IGM ionization process including overlapping of bubbles, a detailed numerical simulation with radiative transfer is needed.  
In the near future, we intend to perform the numerical simulation,
considering the reionization process.


%
\section*{acknowledgments}
HT is supported by the Belgian Federal Office for Scientific,
Technical and Cultural Affairs through the Interuniversity Attraction Pole P6/11.
NS is supported by Grand-in-Aid for Scientific Research No.~22340056
and 18072004.
This research has also been supported in part by World
Premier International Research Center Initiative, MEXT,
Japan.
%

\end{document}